\documentclass[prb,twocolumn,showpacs,floatfix]{revtex4}
\usepackage{graphics,epsfig,amsfonts,amssymb,amsmath}
\begin{document}
\title{Magnetic interactions in iron superconductors studied with a five-orbital model within the
Hartree-Fock and Heisenberg approximations}
\author{M.J. Calder\'on}
\affiliation{Instituto de Ciencia de Materiales de Madrid,
ICMM-CSIC, Cantoblanco, E-28049 Madrid (Spain).
}
\author{G. Le\'on}
\affiliation{Instituto de Ciencia de Materiales de Madrid,
ICMM-CSIC, Cantoblanco, E-28049 Madrid (Spain).
}
\author{B. Valenzuela}
\affiliation{Instituto de Ciencia de Materiales de Madrid,
ICMM-CSIC, Cantoblanco, E-28049 Madrid (Spain).
}
\author{E. Bascones}
\email{leni@icmm.csic.es}
\affiliation{Instituto de Ciencia de Materiales de Madrid,
ICMM-CSIC, Cantoblanco, E-28049 Madrid (Spain).
}

\date{\today}
\begin{abstract}
We have
analyzed the magnetic interactions of a
five orbital model for iron superconductors 
treated both within Hartree-Fock and Heisenberg
approximations. We have found that 
the exchange constants depend non-trivially on the Fe-As-Fe angle and on the charge
and orbital filling. Within the localized picture, columnar ordering is found for
intermediate Hund's coupling $J_H$. At smaller $J_H$, an unusual orbital
reorganization stabilizes checkerboard ordering. Ferromagnetism appears at large $J_H$. 
Ferromagnetic correlations 
are enhanced with electron doping while large hole doping stabilizes checkerboard antiferromagnetism, explaining the change in magnetic interactions upon substitution of Fe by Co or Mn.
 For intermediate and large values of $U$, Hartree-Fock shows similar results as strong coupling though with a double stripe phase instead of ferromagnetism. Itinerancy enhances the stability of the columnar ordering. Comparison of the two approaches reveals a metallic region of the phase diagram where strong coupling physics is determinant. 
\end{abstract}
\maketitle
\section{Introduction}
The origin of magnetism in iron superconductors, believed to be key to
explain their high-T$_c$ superconductivity, is still unsettled. 
Most iron pnictides order with ${(\pi,0)}$ momentum, 
antiferromagnetically (AF) in the $x$-direction and 
ferromagnetically (FM) in the $y$-direction.~\cite{cruz08,zhao08}
The itinerant versus strong coupling origin of
magnetism is at the heart of the debate, with some authors proposing the
coexistence of localized and itinerant moments.~\cite{Castro-Neto,siNJP09,yin10}   
In the itinerant picture the electrons close to the Fermi surface drive the
  ordering through approximate nesting.~\cite{mazin08-2,raghu08,chubukov08,cvetkovic09}  In the strong coupling limit localized spins interact AF. Classically, a ${(\pi,0)}$ state arises when $J_2>J_1/2$, with $J_1$ and $J_2$ the first and second nearest neighbor exchange parameters. A large $J_2$ was justified by the
As-mediated exchange between Fe atoms. 
These, apparently opposite, views of magnetism are minimal descriptions of a 
more complex problem which includes the kinetic energy, the orbital character 
of the electronic bands, and the interactions between the electrons. In iron pnictides interactions are believed to be intermediate
between both limits.~\cite{qazilbash09}  

 The applicability of strong coupling relies on $J_2>J_1/2$. However, 
little is known about the value of the exchange constants. 
Estimates for $J_1$ and $J_2$ have been restricted to a few ab-initio
calculations for specific 
compounds,~\cite{haule09,maPRL09,moonPRB09,moonPRL10,yan10} and
the extraction of the exchange parameters from neutron experiments 
is still controversial.~\cite{zhaonatphys09,wysocki11,strong-coupling-hu11}
The relevance of longer range interactions has also been discussed.\cite{yildirim09}

The situation is even more complex in
FeTe which orders FM along 
one of the diagonals and AF along the other, in a double stripe (DS) 
pattern.  No nesting features compatible with this ordering have been observed.~\cite{xiaPRL09}  A large exchange interaction to third nearest
neighbors $J_3$ has to be introduced to explain the stability of this 
state within a localized picture.~\cite{maPRL09}  
 Moreover, an unexpected FM $J_1$ has  been recently proposed to fit the 
spin-wave spectrum of iron chalcogenides.~\cite{strong-coupling-hu11} 
Finally, the importance of Hund's coupling has also been emphasized in the literature~\cite{hansmannPRL10,haule09,yin11,johannes09} though it is not clear at present how Hund's coupling affects magnetism. 

 In order to connect the itinerant and localized pictures,
we here analyze the magnetic interactions of iron superconductors on
the basis of a five orbital model~\cite{nosotrasprb09} treated both within 
HF and Heisenberg approximations.   Within the strong coupling picture, we find that 
$J_1$ and $J_2$ have a {\it non-trivial}
dependence on the atomic configuration and the Fe-As-Fe angle,  and may
  become ferromagnetic at large Hund's coupling $J_H$. 
For undoped compounds, intermediate values of $J_H$ stabilize the $(\pi,0)$ state, and the generally assumed relation $J_2>J_1/2$ is fulfilled.
$(\pi,\pi)$ 
checkerboard, with $J_2<J_1/2$,  is found for low $J_H$, 
 while FM appears for high $J_H$. 
 The checkerboard ordering at small $J_H$ is stabilized by an
  unusual orbital reorganization and results in an
unexpected sensitivity of the ground state to crystal field parameters. 
 The tendency towards FM, due to virtual transitions involving filled $3z^2-r^2$ or $x^2-y^2$
orbitals, is
enhanced with electron doping, while hole doping stabilizes
checkerboard AF. This can explain the different magnetic orderings observed in the checkerboard antiferromagnetic BaMn$_2$As$_2$, \cite{singh_mn09} with $5$ electrons per Mn, and the ferromagnetic LaOCoAs, \cite{yanagi08} with $7$ electrons per Co.
The HF description reproduces the $(\pi,\pi)-(\pi,0)$ transition with
increasing $J_H$ for intraorbital interaction $U \gtrsim 2.2$ eV allowing us to identify the
metallic regions of the phase diagram where strong coupling physics is relevant. A
DS state shows up at large $U$ and $J_H$, suggesting that it appears as a compromise between kinetic energy cost and the FM tendencies of localized spins. 

\begin{figure*}
\leavevmode
\includegraphics[clip,width=0.95\textwidth]{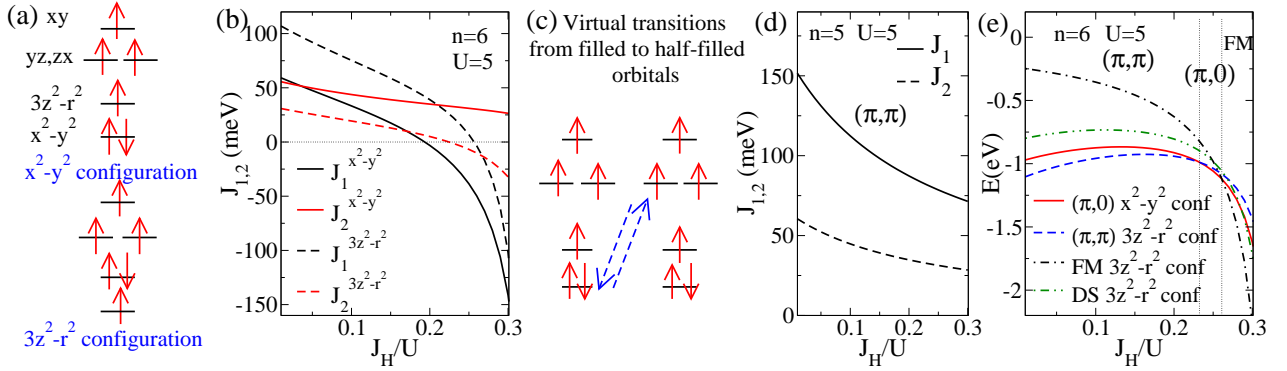}
\vskip -0.4cm
\caption{
(Color online)
(a) Sketch of the two different orbital configurations ($\{\lambda\}=x^2-y^2$ and $\{\lambda\}=3z^2-r^2$) of the $S=2$ state for $n=6$. 
(b) Calculated exchange constants for $n=6$, $U=5$ eV, and $\alpha=35.3^o$ as a function of
Hund's coupling.   (c) Illustration of the virtual transitions involving filled orbitals responsible for the FM exchanges for $n=6$ at large $J_H/U$. (d) Exchange constants as a function of $J_H/U$ for $n=5$ and $U=5$ eV assuming a $S=5/2$ state. A $(\pi,\pi)$ ground state is always favored.
(e) Energies of the magnetic ground states as a function
of $J_H/U$ for the same parameters as in (b). The double stripe DS order, although never the ground state, is
included for comparison. Vertical lines separate regions with different ground states [$(\pi,\pi), (\pi,0)$ and FM].}
\label{fig:exchange}
\end{figure*} 

\section{Model}
The 
Hamiltonian includes intraorbital $U$, interorbital $U'$, 
Hund's coupling $J_H$, and pair hopping $J'$ terms.
\begin{eqnarray}
\nonumber
& H &  = \sum_{i,j,\gamma,\beta,\sigma}t^{\gamma,\beta}_{i,j}c^\dagger_{i,\gamma,\sigma}c_{j,\beta,\sigma}+h.c. 
+ U\sum_{j,\gamma}n_{j,\gamma,\uparrow}n_{j,\gamma,\downarrow}
\\ \nonumber & +&  (U'-\frac{J_H}{2})\sum_{j,\gamma>\beta,\sigma,\tilde{\sigma}}n_{j,\gamma,\sigma}n_{j,\beta,\tilde{\sigma}}
-2J_H\sum_{j,\gamma >\beta}\vec{S}_{j,\gamma}\vec{S}_{j,\beta}
\\
& + &  J'\sum_{j,\gamma\neq
  \beta}c^\dagger_{j,\gamma,\uparrow}c^\dagger_{j,\gamma,\downarrow}c_{j,\beta,\downarrow}c_{j,\beta,\uparrow}
+ \sum_{j,\gamma,\sigma}\epsilon_\gamma n_{j,\gamma,\sigma} \,.
\label{eq:hamiltoniano}
\end{eqnarray}
$i,j$ label the Fe sites in the Fe unit cell, $\sigma$ 
the spin and $\gamma$ and $\beta$ the
five Fe d-orbitals $yz$, $zx$, $xy$, $3z^2-r^2$ and $x^2-y^2$, with $x$ and $y$ axis along the Fe-Fe bonds. We use
$U'=U-2J_H$,~\cite{castellani78} and $J'=J_H$, leaving only two independent interaction parameters, $U$ and $J_H$.
The hopping amplitudes, restricted to first and second neighbors, 
depend on the angle $\alpha$ formed by the Fe-As bonds and the 
Fe-plane.\cite{nosotrasprb09}
$\alpha=35.3^o$, corresponding to the regular Fe-As
tetrahedra, is assumed except when indicated. We take~\cite{nosotrasprb09} $\epsilon_{xy}=0.02$, $\epsilon_{zx,yz}=0$,
$\epsilon_{3z^2-r^2}=-0.55$ and $\epsilon_{x^2-y^2}=-0.6$ for the crystal field.
Energies are  in units of
$(pd\sigma)^2/|\epsilon_d-\epsilon_p|\sim 1$ eV, with $pd\sigma$ the $\sigma$
overlap between the Fe-d and As-p orbitals and 
$|\epsilon_d-\epsilon_p|$ their energy difference.\cite{nosotrasprb09}
For details on the mapping to a classical Heisenberg model and HF
treatment see Appendix \ref{app:A}.

To second order in perturbation theory, starting from localized atomic states,
the 5-orbital Hamiltonian is mapped onto a classical Heisenberg Hamiltonian:
\begin{equation}
E_0^{\{\lambda\}}+{{J^{\{\lambda\}}_1}\over{|S|^2}}\sum_{\langle
  i,j\rangle}\vec{S_i}\vec{S_j}+{{J^{\{\lambda\}}_2}\over{|S|^2}}\sum_{\langle \langle i,j
  \rangle \rangle}\vec{S_i}\vec{S_j}
\label{eq:heisenberg}
\end{equation}
with $\vec{S}_{i}=\sum_{\beta}\vec{S}_{i,\beta}$ the atomic moment and
$\langle i,j \rangle$ and $\langle \langle i,j \rangle \rangle $ restricted 
to first and second nearest neighbors 
respectively. Note that a bicuadratic term $\sim K (S_i S_j)^2$ has been discussed phenomenologically in connection with the structural transition, nematicity, and to reproduce the neutron spectra.\cite{wysocki11,sachdev08,kivelson08,strong-coupling-hu11} The prefactor $K$ would appear to higher order in $t/U$ in perturbation theory and is beyond the scope of this work. Similarly, longer range interactions\cite{yildirim09} are neglected.

To calculate the exchange constants $J_{1,2}^{\{\lambda\}}$ we focus on the largest possible spin state, with each orbital being 
half-filled or filled. For $n=6$, as in undoped 
compounds, this corresponds to $S=2$. This large spin state dominates the
$(\pi,0)$  mean field phase diagram at large $U$.~\cite{nosotrasprl10,nosotrasprl10_2} Due to the small crystal field splitting, we consider two possible atomic 
configurations, labelled by $\{\lambda\}$, for the $S=2$ state with filled $x^2-y^2$ or $3z^2-r^2$ orbitals, see  
Fig.~\ref{fig:exchange} (a). 

\section{Results}

As shown in Fig.~\ref{fig:exchange}(b) 
$J_1^{\{\lambda\}}$ and $J_2^{\{\lambda\}}$  
decrease monotonically with Hund's coupling. 
 Decreasing exchange constants with $J_H$ are generally expected,
but the decrease we find is notably steep, mainly for $J_1$ and at large $J_H/U$, where a change in the
slope happens and the exchange constants even become FM. 
This strong dependence at large $J_H$ is due to virtual transitions from a filled orbital to a
half-filled one on a neighbor atom, see 
Fig.~\ref{fig:exchange}(c) and Appendix \ref{app:A}.
These transitions are favored by the small crystal field
splitting characteristic of iron pnictides. 

The exchange constants and 
their $J_1/J_2$ ratio are very different 
in both atomic configurations.
For $n=6$, in the $x^2-y^2$
configuration favored by the crystal field splitting, $J_2^{x^2-y^2}> 
J_1^{x^2-y^2}/2$ and the $(\pi,0)$ state
is lowest in energy, except at large $J_H/U$ for which FM is expected.  
When comparing the energies between the ordered states in both orbital 
configurations in Fig.~\ref{fig:exchange}(e), the $(\pi,\pi)$ order with 
filled $3z^2-r^2$ becomes the ground state in a wide range of parameters.  The
selection of $(\pi,\pi)$ in the $3z^2-r^2$ configuration is in accordance with 
the exchange constants ratio $J_2^{3z^2-r^2}<J_1^{3z^2-r^2}/2$. Remarkably, 
the gain in magnetic energy, associated to the large value of the direct hopping $t_{x^2-y^2,x^2-y^2}$,
\cite{nosotrasprb09}
compensates for the cost in crystal field.
Due to the small energy
difference between the two states, very small changes in the crystal field can stabilize $(\pi,0)$ 
for smaller values of $J_H$, see Appendix \ref{app:B}.

Note that  
the exchange constants in the 
$(\pi,0)$ and $(\pi,\pi)$ states may be different. Thus 
estimating these constants by comparing the energy of the different magnetic states with those predicted by a mapping to a
Heisenberg model~\cite{maPRL09,moonPRB09,moonPRL10,yan10} with the same $J_1$ and $J_2$ 
may lead to errors. 
Moreover, our results indicate that the filled orbitals are not inert
for magnetism, so the use of 4-orbital models  which neglect them is questionable.~\cite{hansmannPRL10,laad2011}

 With electron doping, the number of filled orbitals increases. For $n=7$,
  $(\pi,0)$ order is found at  small $J_H$ and FM behavior appears at a
  smaller value of $J_H$ compared to $n=6$ (not shown).
On the other hand, there are no filled orbitals at $n=5$ in its highest 
spin state ($S=5/2$), implying a weaker dependence on $J_H/U$ of the exchange 
parameters, always AF, see Fig.~\ref{fig:exchange}(d).
In this case $J_2<J_1/2$ and $(\pi,\pi)$ ordering is found for all $J_H$. Therefore,  a clear
asymmetry is found in the magnetic interactions with strong electron-hole
doping. This is consistent with experimental observations: BaMn$_2$As$_2$ and 
LaOCoAs show checkerboard ordering and ferromagnetism, respectively.~\cite{yanagi08,singh_mn09}    

\begin{figure}
\leavevmode
\includegraphics[clip,width=0.42\textwidth]{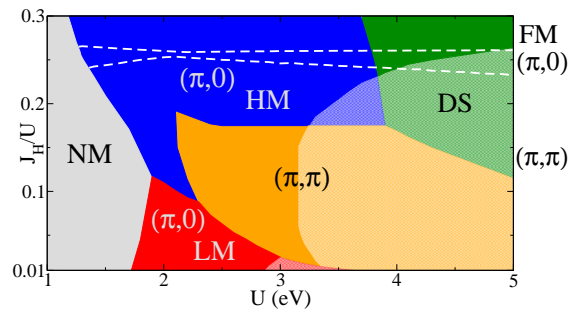}
\vskip -0.4cm
\caption{
(Color online) Hartree Fock magnetic phase diagram calculated at fixed density
$n=6$ as a function of $U$ and $J_H/U$. Grey stands for non-magnetic (NM), blue for the AF $(\pi,0)$ high moment (HM) state
satisfying Hund's rule, red for a the AF $(\pi,0)$ low moment (LM) state which
violates Hund's rule, orange for
AF $(\pi,\pi)$ and green for the double stripe (DS) state with charge modulation (see text). Shaded areas mark insulating phases. Superposed dashed lines on the magnetic regions show the phase transition lines between the $(\pi,\pi)$ and HM $(\pi,0)$ states and between
the HM $(\pi,0)$ and the FM phase predicted by the $S=2$ Heisenberg 
model.}
\label{fig:diagfasenfijo}
\end{figure}

A $(\pi,\pi)-(\pi,0)$ transition with increasing $J_H$ is also present in
  the HF phase diagram in Fig.~\ref{fig:diagfasenfijo}. Crystal field 
  sensitivity and orbital
  reorganization, similar to that found in the localized picture, is realized, 
see Appendix \ref{app:B}. The transition between $(\pi,\pi)$  and
$(\pi,0)$ 
is accompanied by a charge transfer between 
$x^2-y^2$ and $3z^2-r^2$. In the $(\pi,0)$ state, $3z^2-r^2$ is emptied while 
$x^2-y^2$ gets filled as $U$ increases.~\cite{nosotrasprl10}
In the $(\pi,\pi)$ state, $3z^2-r^2$ fills 
with increasing $U$ while the other orbitals tend to
half filling. The $(\pi,0)$ state becomes more stable if the splitting between
$x^2-y^2$ and $3z^2-r^2$ increases, see Appendix \ref{app:B}.

As previously found,~\cite{nosotrasprl10} 
two different $(\pi,0)$ states show
up at the HF level. In the LM state, opposite orbital magnetizations
within the same atom result in a low magnetic
moment which violates Hund's rule. This state is stabilized thanks to the
anisotropy of the orbital exchange constants~\cite{nosotrasprl10} and has been
proposed to explain the low magnetic moment found
experimentally.~\cite{nosotrasprl10,cricchio09,liu11} In the HM state, all
the orbital magnetizations point in the same direction. The strong
coupling predictions, dashed lines in Fig.~\ref{fig:diagfasenfijo}, 
are valid for comparison with this HM state. 

As expected within a weak coupling description, and opposite to the strong
  coupling predictions, for small values of the 
interaction a HM $(\pi,0)$ is found close to the non-magnetic boundary. 
In this region nesting seems to stabilize the $(\pi,0)$ ordering, although the
electronic reconstruction happens not only at the Fermi surface but also at
higher energies. Similar physics has been discussed within density functional theory
calculations.\cite{johannes09} 
On the other hand, the stability of the $(\pi,\pi)$ state for 
$U \gtrsim 2.2$ eV, 
the orbital reorganization at the
  transition, and the strong crystal field sensitivity cannot be understood
  within the nesting picture and are a clear signature of localized physics. 
  Note that this is found even for metallic states. Even for $U \gtrsim 2.2$ eV
the $(\pi,0)$ state is more prominent in the HF phase diagram than in the
Heisenberg description. This suggests some influence of itinerancy in
  stabilizing the $(\pi,0)$ state.

\begin{figure}
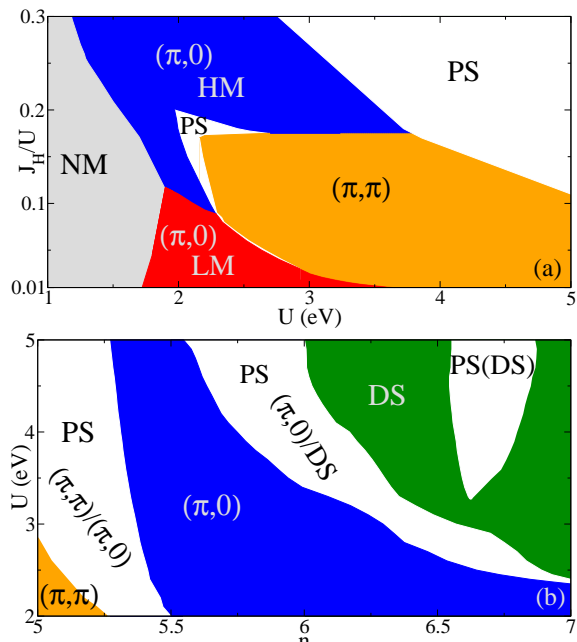

\leavevmode
\includegraphics[clip,width=0.42\textwidth]{PDn6-HF-PS3.eps}
\includegraphics[clip,width=0.42\textwidth]{diagfaseJp22color.eps}
\vskip -0.4cm
\caption{
(Color online)  Mean field magnetic phase diagram in the grand canonical ensemble as a function of $U$ and $J_H/U$ and for $n=6$ (a) and as a function of $U$ and $n$ for
$J_H/U=0.22$ (b). Same color code as in Fig.~(\ref{fig:diagfasenfijo})
applies. White regions are unstable towards phase separation.}
\label{fig:diagfasemufijon6}
\end{figure}

At large $J_H$ the kinetic energy effects present at the HF level prevent the 
FM solution 
found in the strong coupling description to arise. Instead, a double stripe 
solution is found at large $U$ and $J_H$. This DS state is charge modulated ($5$ or $7$ electrons for the $n=6$ case) in a checkerboard fashion. 
Note that in the Heisenberg picture, a homogeneous DS state, though not the ground state, is lower in energy than both $(\pi,\pi)$ and $(\pi,0)$ at large $J_H/U$, see Fig.~\ref{fig:exchange}(e). 
The homogeneous DS state, with $n$ electrons in each atom, is difficult to stabilize in the HF calculations, mainly at large $J_H$. When stabilized, it has larger energy than the charge modulated DS
state. As shown in Fig.~\ref{fig:diagfasemufijon6}, the DS state is unstable towards phase separation at $n=6$, see Appendix \ref{app:A}, but can be stabilized by electron doping. 
As expected for first order transitions, PS appears at the boundary between the different magnetic phases. The instability at $n \gtrsim 6.5$, which only involves the DS state, has a different origin: it is caused by a negative compressiblity related to the charge modulation.

Fig.~\ref{fig:diagfasemufijon6}(b) also evidences an 
electron-hole asymmetry around the undoped composition $n=6$. In agreement with the strong coupling  
predictions, for $n=5$ $(\pi,\pi)$ correlations dominate, 
even at large $J_H$ (calculations at $J_H/U=0.28$ give 
similar results). When $n$ increases, the $(\pi,0)$ state becomes lower 
in energy, being the ground state in a wide region of parameters.   
For $n>6$ and intermediate or large $U$ the DS state is
found. A larger tendency to DS solutions with increasing $n$ is also expected in
the Heisenberg description.  Within HF, the electron-hole 
asymmetry close to $n=6$ gets reduced as $U$ decreases.  

\begin{figure}
\leavevmode
\includegraphics[clip,width=0.42\textwidth]{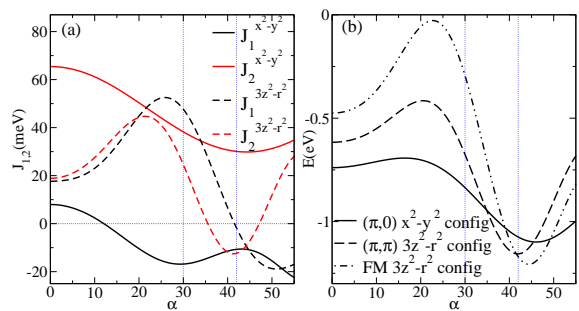}
\vskip -0.4cm
\caption{
(Color online)  (a) First and second nearest neighbor exchange constants
$J_1^{\{\lambda\}}$ and $J_2^{\{\lambda\}}$ as a function of $\alpha$, the angle
formed by the Fe-As bond and the Fe plane, for $n=6$, $U=5$ eV, and
$J_H/U=0.22$. (b) Energies of the most competitive magnetic orderings as a
function of $\alpha$ for the parameters in (a). The sequence of magnetic
orderings with increasing $\alpha$ is $(\pi,0)$, $(\pi,\pi)$, and FM. Vertical
lines delimitate the
experimentally relevant values of $\alpha$.}
\label{fig:angle} 
\end{figure}

We finally analyze the effect of the Fe-As-Fe geometry on the magnetic
interactions. A direct relation between the Fe-As-Fe angle and the critical
superconducting temperature~\cite{iyo08,zhao08,kuroki09-2} as well as on the magnetic 
ordering~\cite{strong-coupling-hu11} has been claimed.
As illustrated in Fig. \ref{fig:angle}, the exchange constants vary 
non-monotonically with $\alpha$ due to the relative importance of the 
hopping (to both first and second neighbors) via the As.~\cite{nosotrasprb09,note-cf}
The dependences of $J_1$ and $J_2$ on $\alpha$ are different for the $x^2-y^2$ and $3z^2-r^2$ configurations, with $J_1/J_2$ changing with $\alpha$.  
Checkerboard and ferromagnetic ordering are
  more stable for elongated tetrahedra.

\section{Summary}
We have analyzed 
the magnetic interactions and ground state of a five
orbital model for iron pnictides by means of Heisenberg and
Hartree Fock approaches. 
We have calculated the exchange constants of the strong coupling model 
and show that their value and sign depend non-trivially on the Fe-As-Fe 
angle, the orbital filling, the number 
of electrons per Fe, and the Hund's coupling $J_H$. 
 A $(\pi,\pi)-(\pi,0)$ transition which involves orbital reorganization
  is present in both approaches.  This cannot be explained within the nesting
  picture and allows us to identify a metallic region of the phase diagram where strong
  coupling physics is relevant. On the contrary the magnetic ordering close to
  the non-magnetic transition could be due to nesting. Both pictures
  seem to be important to describe magnetism in iron pnictides. 
Our results uncover the effect of $J_H$ on the stabilization of $(\pi,0)$ ordering. Due to virtual transitions involving filled orbitals, a large $J_H$ produces a 
strong decrease of the exchange constants leading to small AF or even  FM 
$J_1$. This could help understand the neutron scattering
experiments.~\cite{strong-coupling-hu11}
  Ferromagnetism
  appears at large $J_H$ in the Heisenberg approach, while a double stripe
  phase shows up in Hartree-Fock.  Electron doping enhances ferromagnetic (double-stripe) 
  tendencies, while large hole-doping leads to checkerboard ordering, in agreement with experiments.~\cite{yanagi08,singh_mn09} 

Note added. Recently, similar results with hole doping were obtained by the variational Monte Carlo method.\cite{imadaPRL2012}

We have benefited from conversations with F. Guinea, A. Chubukov and N. Perkins. We acknowledge funding from Ministerio de Ciencia e Innovaci\'on through Grants No. FIS 2008-00124, FIS 2009-08744 and Ram\'on y 
Cajal contract, and from CSIC through Grants No. PIE-200960I033 and PIE-200960I180.

\appendix
\section{Calculation details}
\label{app:A}

Starting from localized atomic states we map the 5-orbital interacting
Hamiltonian of Eq.(1) in a classical Heisenberg 
Hamiltonian with exchange interactions to first and second nearest neighbors

\begin{equation}
E_0^{\{\lambda\}}+\frac{J^{\{\lambda\}}_1}{|S|^2}\sum_{\langle
  i,j\rangle}\vec{S_i}\vec{S_j}+\frac{J^{\{\lambda\}}_2}{|S|^2}\sum_{\langle
  \langle i,j\rangle \rangle}\vec{S_i}\vec{S_j}  \,\, .
\label{eq:heisenbergsupp}
\end{equation}   

Here $\{\lambda\}$ labels the considered orbital configuration (as defined in Fig. 1 (a)),  $\vec{S}_{i}=\sum_{\beta}\vec{S}_{i,\beta}$ the atomic moment and
$\langle i,j \rangle$ and $\langle \langle i,j \rangle \rangle $ are restricted to first and second nearest
neighbors respectively. We restrict to atomic states with maximum total spin
$S$ ($S=2$ for $n=6$ and $S=5/2$ for $n=5$) and maximum $|S_z|$ ($S_z=\pm 2$
and $S_z=\pm 5/2$ for $n=6$ and $n=5$ respectively). With this, $E_0^{\{\lambda\}}$ and $J^{\{\lambda\}}_{1,2}$ are

\begin{equation}
E_0^{\{\lambda\}}=\sum_{\langle i,j \rangle, \langle\langle i,j \rangle \rangle}\frac{E^{\{\lambda\}}_{P;i,j}+E^{\{\lambda\}}_{AP;i,j}}{4} \,\, ,
\label{eq:e0classical}
\end{equation}
\begin{equation}
J^{\{\lambda\}}_{1,2}=\frac{E^{\{\lambda\}}_{P;i,j}-E^{\{\lambda\}}_{AP;i,j}}{4} \,\, ,
\label{eq:j1classical}
\end{equation}
with $i,j$ two 
first (second) nearest neighbors for $J_1$ ($J_2$). 
$E^{\{\lambda\}}_{P;i,j}$ and $E^{\{\lambda\}}_{AP;i,j}$ are the energies corresponding to two parallel or antiparallel spins at $i$ and $j$.
The factor of $4$ (instead of $2$) corrects the bond double-counting in 
Eq.(\ref{eq:heisenbergsupp}). 
When using Eqs.~(\ref{eq:e0classical}) and (\ref{eq:j1classical})  we are 
neglecting quantum fluctuations. In this sense we are treating
the spin classically. As discussed in the text and below, the classical
approximation provides good understanding on the magnetic interactions
dominant in iron pnictides. Quantum fluctuations are expected to be of little 
importante for large spins. The spin states $S=2$ and $S=5/2$ discussed in the
text are large enough to justify the classical treatment of spins.   

P and AP energies can be written in terms of atomic and magnetic energies
\begin{equation}
E_{P,AP;i,j}^{\{\lambda\}}=E^{\{\lambda\}}_{at}+
E^{\{\lambda\},mag}_{P,AP;i,j} \,\, ,
\end{equation} 
where $E^{\{\lambda\}}_{at}$ accounts for the
interaction and crystal field energy to charge the atoms to the selected $n=6$ or $n=5$
states, while $E^{\{\lambda\},mag}_{P,AP;i,j}$ is the magnetic contribution. 
$E^{\{\lambda\}}_{at}$
enters in the value of the constant $E^{\{\lambda\}}_0$ but cancels out in the
expression of the exchange constants $J^{\{\lambda\}}_{1,2}$.
We simplify $E^{\{\lambda\}}_{at}$ to $E^{x^2-y^2}_{at}=0$ and
$E^{3z^2-r^2}_{at}=\epsilon_{3z^2-r^2}-\epsilon_{x^2-y^2}$ per atom, as
the other terms drop out from the calculation.  For $n=6$ 

\begin{eqnarray}
E^{\{\lambda\},mag}_{AP;i,j} = -\sum_{\eta,\nu}\frac{2(t^{\nu,\eta}_{i,j})^2}{J'^2+
  \Delta E_{-}^2}\left(\frac{\Delta E_{-}^2}{U+3
    J_H+\epsilon_\eta-\epsilon_\nu+\Delta
    E_+}  \nonumber \right.
\\ 
\left.
+  \frac{J'^2}{U+3J_H+\epsilon_\eta-\epsilon_\nu+\Delta E_-} \right) \nonumber
\\
-\sum_{\nu}2(t^{\lambda,\nu}_{i,j})^2\left
   (\frac{1/5}{U-3J_H+\epsilon_\eta-\epsilon_\lambda} +\frac{4/5}{U+2J_H+\epsilon_\eta-\epsilon_\lambda}\right) 
\label{eq:e_af}
\end{eqnarray}
Here $\eta$ and $\nu$ label half-filled orbitals and $\lambda$ refers to the
filled one; $t^{\beta, \gamma}_{i,j,}$ are the hopping amplitudes between
orbitals $\beta$ and $\gamma$ in atoms $i$ and $j$ respectively.\cite{nosotrasprb09} $\Delta E_{\pm}=\epsilon_\nu-\epsilon_{\lambda} \pm \left((\epsilon_{\lambda}-\epsilon_\nu)^{2}+J'^2 \right
    )^{1/2}$. The first term in Eq.~(\ref{eq:e_af}) includes virtual transition from
    half-filled orbitals in one atom to half-filled orbitals in a neighbour
    atom and takes into account that the intermediate state is not
    an eigenstate of the pair hopping operator. The second term is associated
    with virtual transitions of an electron from the filled orbital to a
    half-filled orbital in the other atom. Only the electron whose spin is
    opposite to the magnetic moment of the neighbor atom can hop. The state 
left behind is not an eigenstate of the Hund's term. 

When the magnetic moments of the two atoms are parallel the only transitions
allowed involve the filled orbital. The contribution of these transitions is 
\begin{equation}
E^{\{\lambda\},mag}_{P;i,j}=-\sum_{\nu}\frac{2(t^{\lambda,\nu}_{i,j})^2}{
  U-3J_H+\epsilon_{\lambda}-\epsilon_\nu} \,\, .
\label{eq:j1}
\end{equation}

For $n=5$ and $S=5/2$, $E^{S=5/2,mag}_{P;i,j}=0$ and 
 \begin{equation}
E^{S=5/2,mag}_{AP;i,j}=-\sum_{\nu \eta}\frac{2(t^{\eta,\nu}_{i,j})^2}{
  U+4J_H+\epsilon_\eta-\epsilon_\nu} \,\, .
\label{eq:j1}
\end{equation}

In this latter case, the exchange constants are clearly AF. Whether a
$(\pi,\pi)$ or $(\pi,0)$ state is favoured depends on the relative values of
$J_1$ and $J_2$. As shown in Fig. 1, for $n=5$ $(\pi,\pi)$ ordering is
always preferred. 

The $n=6$ case is more interesting. For $J_H=0$ the contribution of the transitions
which depart from filled orbitals is the same with P or AP
ordering. Consequently they do not affect the magnetic ordering, and the 
exchange
constants, driven by the transitions from half-filled orbitals are AF. 
With increasing $J_H$ their contribution is larger in the P case. At a given 
$J_H$ these transitions favor FM ordering. For small $J_H$ the transitions
from half-filled orbitals still dominate, but at large
$J_H$ $J^{\{\lambda\}}_1$ and $J^{\{\lambda\}}_2$ become ferromagnetic. 
Due to different $t^{\gamma, \beta}_{i,j}$ in each of these terms the
value of $J_H$ at which this happens is different for each exchange constant.

\begin{figure}
\leavevmode
\includegraphics[clip,width=0.5\textwidth]{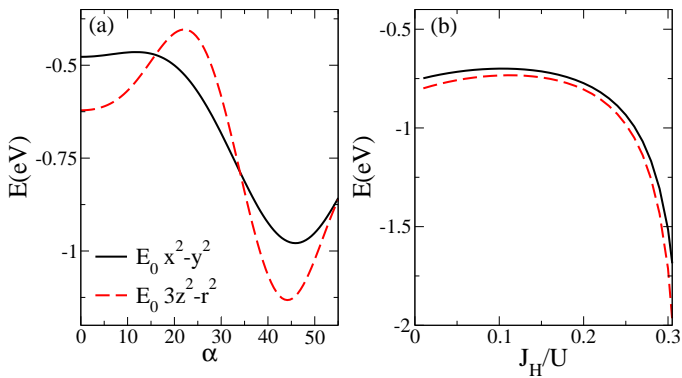}
\caption{
(Color online) Dependence of $E^{x^2-y^2}_0$ and $E^{3z^2-r^2}_0$
for $n=6$ and $U=5$ eV as a function of the Fe-As-Fe angle $\alpha$ for $J_H/U=0.22$ (a) and Hund's coupling for $\alpha=35.3^o$(b).}
\label{fig:e0supplementary}
\end{figure}  

Another interesting aspect regards $E^{\{\lambda\}}_0$. 
The contribution of $E^{\{\lambda\},mag}_{P,AF;i,j}$ to $E^{\{\lambda\}}_0$
can be seen as the energy gain associated to creating magnetic moments, even
if on average they do not order. 
When mapping to Heisenberg models it is usual to disregard $E_0$ because 
if a single atomic state is selected its
value does not affect the differences in energy between magnetic states. 
However,  for $n=6$, the small crystal field splitting between $x^2-y^2$ and
$3z^2-r^2$ requires the inclusion of two different atomic states. Due to different
hopping amplitudes involving these orbitals, $E^{x^2-y^2}_0 \neq E^{3z^2-r^2}_0$ is found.
This can be observed in Fig.~\ref{fig:e0supplementary} as a function of the
Fe-As-Fe and Hund's coupling.  
Consequently $E^{\{\lambda\}}_0$ helps select the atomic state
$\{\lambda\}$, with the a priori unexpected result that the atomic state in
which $3z^2-r^2$ is filled is preferred in a wide range of parameters. 
Note that the formation of magnetic moments is favoured in the experimentally
relevant range of Fe-As-Fe angle and at large Hund's coupling.  
The large values of $E^{\{\lambda\}}_0$ also explain the large values 
of the total
energy, despite the relatively small values of the exchange constants.

We have compared these predictions with a self-consistent mean-field Hartree Fock
calculation which includes non-magnetic (NM),  FM, and AF states 
with $Q=(\pi,0)$, $Q=(\pi,\pi)$,  and DS ordering.
In the DS calculation the system is divided into two interpenetrating 
lattices coupled via first nearest neighbor hopping terms. 
The axis and orbital basis is rotated and a $(\pi,0)$ state 
along the Fe-diagonals is assumed. In the mean-field calculation
only spin and orbital-diagonal average terms are kept: 

\begin{eqnarray}
n_{\gamma}&=&\sum_{k,\sigma} \langle c^\dagger_{k,\gamma,\sigma}
c_{k,\gamma,\sigma}\rangle 
\\ \nonumber
m_{\gamma}&=&\sum_{k} \left [\langle c^\dagger_{k+\bf{Q},\gamma,\uparrow}
c_{k,\gamma,\uparrow}\rangle- \langle c^\dagger_{k+\bf{Q},\gamma,\downarrow}
c_{k,\gamma,\downarrow}\rangle\right ] .
\label{eq:meanfield}
\end{eqnarray}

We have checked that this approximation does not have any influence on the
mean field results: Disregarded terms are of order $\sim 10^{-4}$.
Several initial configurations for $n_\gamma$ and $m_\gamma$ are probed for
each ordered state. A self-consistent solution is considered to be ordered
(insulating) when the magnetic moment $m$ (gap) is larger than $0.001$. 

The phase diagrams in Fig.~2 and Fig.~\ref{fig:supplementary3} (right panel) have
been obtained comparing the energies of 
the mean-field solutions at fixed density. 
Different phases have a different energy $E$ versus density $n$ 
relation. Jumps
in the chemical potential $\mu$ appear at first-order boundaries between 
ordered phases. In such situations phase separation is expected. In a range of
densities the system decreases its energy by allowing relative fractions of
the two neighboring magnetic phases. $n=6$ may belong to this range of
densities for values of the interactions close to phase boundaries. To
determine the region of phase space unstable towards phase separation in Fig.~3 we work
in the grand-canonical ensemble (fixed chemical potential) and compare $E- \mu
n$ for the different states. The chemical potential can be calculated as $\partial E/\partial n$ or by
looking at the energy of the last state occupied in metallic systems, both
methods giving the same value. $\mu$ jumps discontinuously when crossing a
gap. According to Maxwell's construction, 
phase boundaries between two phases $1$ and $2$
are given by $E_1-\mu n_1 =E_2-\mu n_2$. Neither phase $1$ nor phase $2$ are
stable in the range $n_1 < n < n_2$. A phase separated mixture of both phases
arises.

Phase separation for a given phase will also appear with negative
compressibility, i.e. $\mu$ decreases with increasing $n$. This is observed
in the DS state in several ranges of $n$, including $n=6$ and $n \leq 6.5$. 
In this case phase
separation happens between states with the same magnetic ordering but
different density. Densities $n_1$ and $n_2$ are determined as above. Note,
that while at $n=6$ DS is intrinsically unstable towards phase separation, at this
density phase separation between DS and $(\pi,0)$ is favored.

\begin{figure}
\leavevmode
\includegraphics[clip,width=0.5\textwidth]{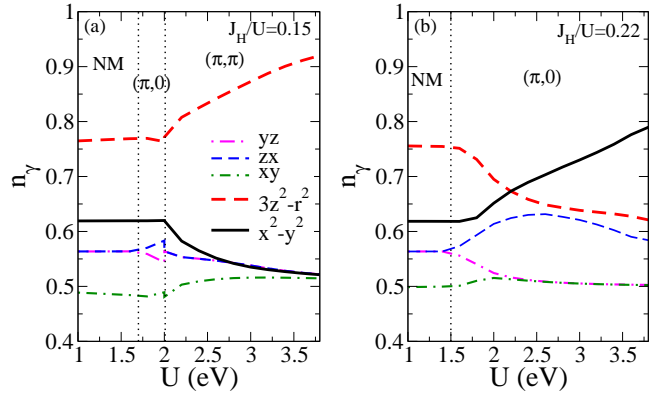}
\caption{
(Color online) (a) and (b) Dependence of the orbital filling as a function of
the intraorbital interaction for $J_H/U=0.15$ and $J_H/U=0.22$
respectively. Vertical lines separate regions with different ground
states.}
\label{fig:supplementary2}
\end{figure} 

\section{$(\pi,0)$ versus $(\pi,\pi)$. Crystal field sensitivity}
\label{app:B}
\begin{figure*}
\leavevmode
\includegraphics[clip,width=0.39\textwidth]{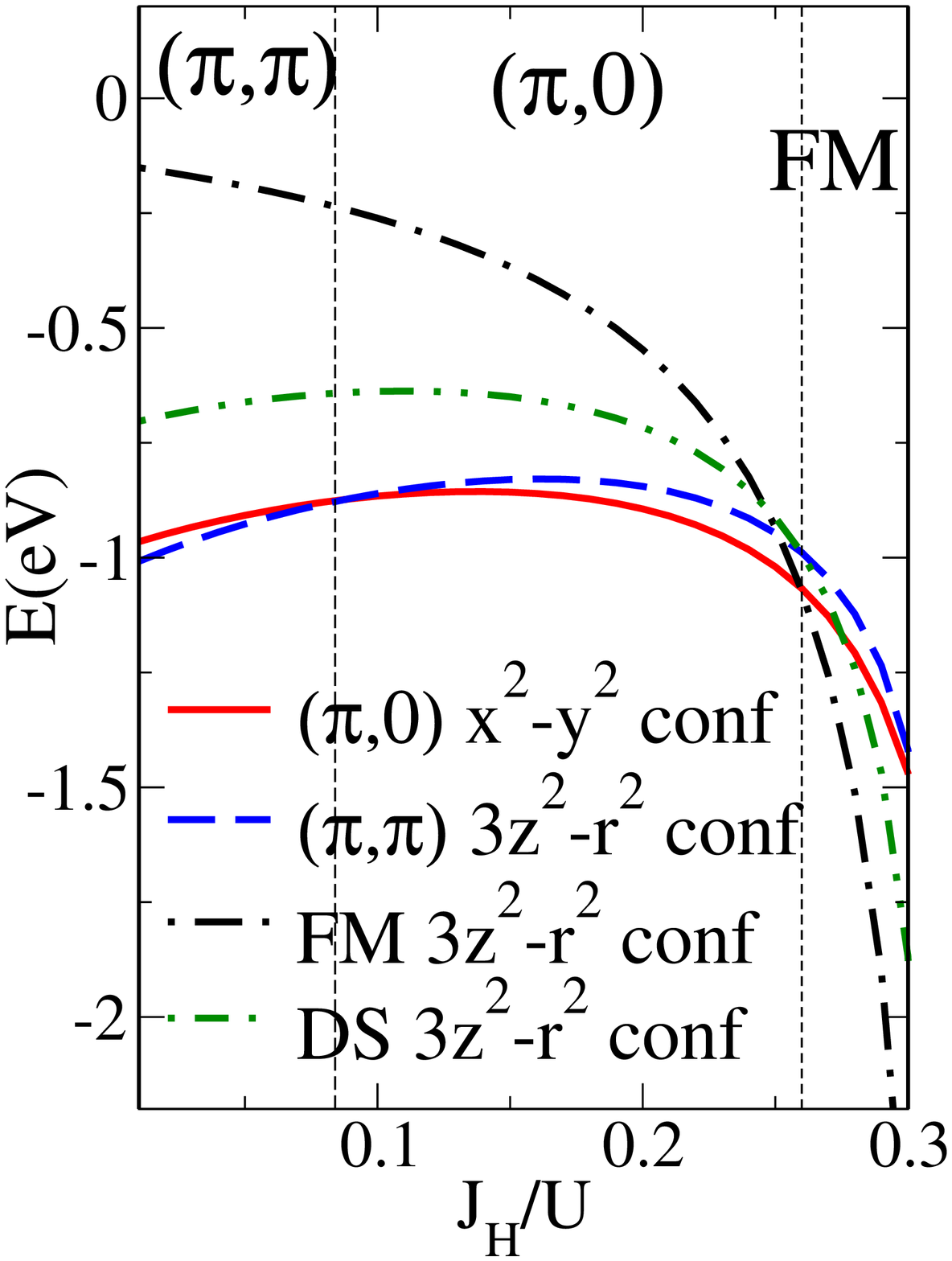}
\includegraphics[clip,width=0.56\textwidth]{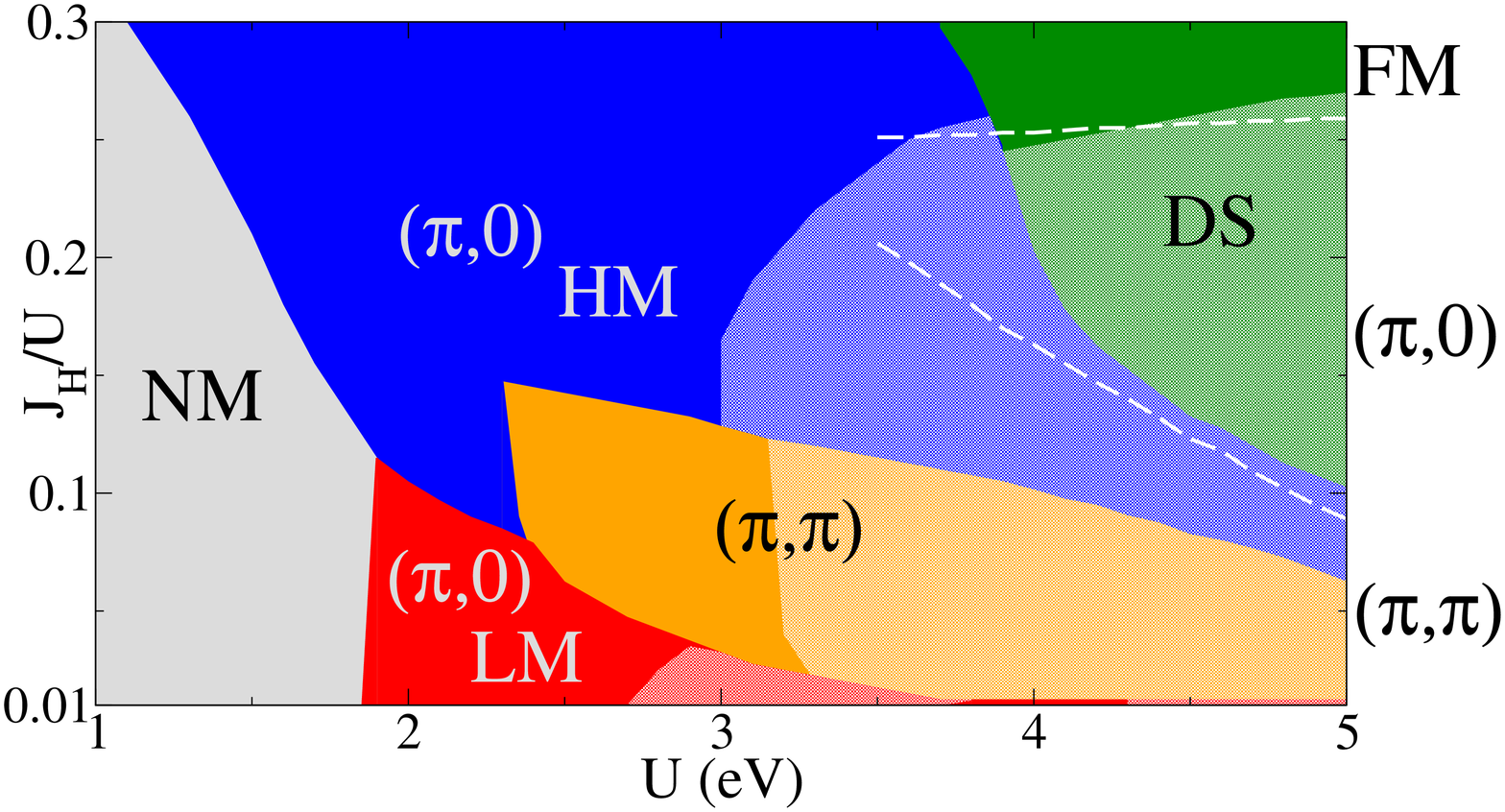}
\caption{
(Color online) (Left) Energies of the magnetic ground states as a function
of $J_H/U$ calculated within the Heisenberg picture and (right) Hartree-Fock phase
diagram for the same parameters as in Fig.~2 and Fig.~3, except for the crystal
field splitting $\epsilon_{3z^2-r^2}-\epsilon_{x^2-y^2}$  which is 100 meV
   larger here. In the right panel the color code is the same as in Fig.~2. Superposed dashed 
lines on the magnetic regions show the phase transition lines between the $(\pi,\pi)$ and HM $(\pi,0)$ states and between
the HM $(\pi,0)$ and the FM phase predicted by the $S=2$ Heisenberg 
model.}
\label{fig:supplementary3}
\end{figure*} 

Fig.~2 shows the phase diagram corresponding to $n=6$ calculated
at fixed density. The
mean field ground state evolves from a non-magnetic solution with zero magnetic
moment and orbital fillings close to the non-interacting ones at small $U$ to
$(\pi,0)$ and $(\pi,\pi)$ states with $m=4 \mu_B$, with $\mu_B$ a Bohr
magneton, as in an atomic $S=2$ state at large $U$. As seen in Fig.~\ref{fig:supplementary2}, in the $(\pi,0)$  state
every orbital except $x^2-y^2$ tends to half  
filling at large $U$, $x^2-y^2$ becoming completely filled. This orbital filling is expected on the basis of the crystal field
 splitting. On the other hand, in the $(\pi,\pi)$ state, it is the $3z^2-r^2$
 orbital the one which fills completely while $x^2-y^2$ tends to
 half-filling. 
This orbital filling competes with the crystal field, in agreement with the
 strong coupling predictions. 

 The competition between
   crystal field and magnetic energy suggests a strong sensitivity of the
   $(\pi,\pi)-(\pi,0)$ transition to the crystal field splitting 
   $\epsilon_{3z^2-r^2}-\epsilon_{x^2-y^2}$. Such a sensitivity is manifest in
   Fig.~\ref{fig:supplementary3} where the stability of the different phases
   is shown for a crystal field splitting 
   $\epsilon_{3z^2-r^2}-\epsilon_{x^2-y^2}=150$ meV,  $100$ meV 
   larger  than the one used in 
   Fig.~2 and Fig.~3.  This modification of the crystal field is below the accuracy of the tight-binding.
   The region of stability of the $(\pi,0)$ state in 
   Fig.~\ref{fig:supplementary3} is considerably larger than in
   Figs.~2 and 3.

    This orbitally reorganized $(\pi,\pi)$ state is a clear
   signature of strong-coupling physics for $U \gtrsim 2.2$ eV. On the
   contrary, its absence for smaller values of $U$ suggests a nesting-driven
   $(\pi,0)$ state.

\bibliography{pnictides}

\begin{thebibliography}{38}
\expandafter\ifx\csname natexlab\endcsname\relax\def\natexlab#1{#1}\fi
\expandafter\ifx\csname bibnamefont\endcsname\relax
  \def\bibnamefont#1{#1}\fi
\expandafter\ifx\csname bibfnamefont\endcsname\relax
  \def\bibfnamefont#1{#1}\fi
\expandafter\ifx\csname citenamefont\endcsname\relax
  \def\citenamefont#1{#1}\fi
\expandafter\ifx\csname url\endcsname\relax
  \def\url#1{\texttt{#1}}\fi
\expandafter\ifx\csname urlprefix\endcsname\relax\def\urlprefix{URL }\fi
\providecommand{\bibinfo}[2]{#2}
\providecommand{\eprint}[2][]{\url{#2}}

\bibitem[{\citenamefont{de~la Cruz et~al.}(2008)\citenamefont{de~la Cruz,
  Huang, Lynn, Li, Ratcliff, Zarestky, Mook, Chen, Luo, Wang et~al.}}]{cruz08}
\bibinfo{author}{\bibfnamefont{C.}~\bibnamefont{de~la Cruz}},
  \bibinfo{author}{\bibfnamefont{Q.}~\bibnamefont{Huang}},
  \bibinfo{author}{\bibfnamefont{J.}~\bibnamefont{Lynn}},
  \bibinfo{author}{\bibfnamefont{J.}~\bibnamefont{Li}},
  \bibinfo{author}{\bibfnamefont{W.}~\bibnamefont{Ratcliff}},
  \bibinfo{author}{\bibfnamefont{J.}~\bibnamefont{Zarestky}},
  \bibinfo{author}{\bibfnamefont{H.}~\bibnamefont{Mook}},
  \bibinfo{author}{\bibfnamefont{G.}~\bibnamefont{Chen}},
  \bibinfo{author}{\bibfnamefont{J.}~\bibnamefont{Luo}},
  \bibinfo{author}{\bibfnamefont{N.}~\bibnamefont{Wang}}, \bibnamefont{et~al.},
  \bibinfo{journal}{Nature} \textbf{\bibinfo{volume}{453}},
  \bibinfo{pages}{899} (\bibinfo{year}{2008}).

\bibitem[{\citenamefont{Zhao et~al.}(2008)\citenamefont{Zhao, Huang, de~la
  Cruz, Li, Lynn, Chen, Green, Chen, Li, Li et~al.}}]{zhao08}
\bibinfo{author}{\bibfnamefont{J.}~\bibnamefont{Zhao}},
  \bibinfo{author}{\bibfnamefont{Q.}~\bibnamefont{Huang}},
  \bibinfo{author}{\bibfnamefont{C.}~\bibnamefont{de~la Cruz}},
  \bibinfo{author}{\bibfnamefont{S.}~\bibnamefont{Li}},
  \bibinfo{author}{\bibfnamefont{J.}~\bibnamefont{Lynn}},
  \bibinfo{author}{\bibfnamefont{Y.}~\bibnamefont{Chen}},
  \bibinfo{author}{\bibfnamefont{M.}~\bibnamefont{Green}},
  \bibinfo{author}{\bibfnamefont{G.}~\bibnamefont{Chen}},
  \bibinfo{author}{\bibfnamefont{G.}~\bibnamefont{Li}},
  \bibinfo{author}{\bibfnamefont{Z.}~\bibnamefont{Li}}, \bibnamefont{et~al.},
  \bibinfo{journal}{Nature Materials} \textbf{\bibinfo{volume}{7}},
  \bibinfo{pages}{953} (\bibinfo{year}{2008}).

\bibitem[{\citenamefont{Wu et~al.}(2008)\citenamefont{Wu, Phillips, and
  Castro~Neto}}]{Castro-Neto}
\bibinfo{author}{\bibfnamefont{J.}~\bibnamefont{Wu}},
  \bibinfo{author}{\bibfnamefont{P.}~\bibnamefont{Phillips}}, \bibnamefont{and}
  \bibinfo{author}{\bibfnamefont{A.~H.} \bibnamefont{Castro~Neto}},
  \bibinfo{journal}{Phys. Rev. Lett.} \textbf{\bibinfo{volume}{101}},
  \bibinfo{pages}{126401} (\bibinfo{year}{2008}).

\bibitem[{\citenamefont{Si et~al.}(2009)\citenamefont{Si, Abrahams, Dai, and
  Zhu}}]{siNJP09}
\bibinfo{author}{\bibfnamefont{Q.}~\bibnamefont{Si}},
  \bibinfo{author}{\bibfnamefont{E.}~\bibnamefont{Abrahams}},
  \bibinfo{author}{\bibfnamefont{J.}~\bibnamefont{Dai}}, \bibnamefont{and}
  \bibinfo{author}{\bibfnamefont{J.-X.} \bibnamefont{Zhu}},
  \bibinfo{journal}{New Journal of Physics} \textbf{\bibinfo{volume}{11}},
  \bibinfo{pages}{045001} (\bibinfo{year}{2009}).

\bibitem[{\citenamefont{Yin et~al.}(2010{\natexlab{a}})\citenamefont{Yin, Lee,
  and Ku}}]{yin10}
\bibinfo{author}{\bibfnamefont{W.-G.} \bibnamefont{Yin}},
  \bibinfo{author}{\bibfnamefont{C.-C.} \bibnamefont{Lee}}, \bibnamefont{and}
  \bibinfo{author}{\bibfnamefont{W.}~\bibnamefont{Ku}},
  \bibinfo{journal}{Physical Review Letters} \textbf{\bibinfo{volume}{105}},
  \bibinfo{pages}{107004} (\bibinfo{year}{2010}{\natexlab{a}}).

\bibitem[{\citenamefont{Mazin et~al.}(2008)\citenamefont{Mazin, Johannes,
  Boeri, and K.~Koepernik}}]{mazin08-2}
\bibinfo{author}{\bibfnamefont{I.}~\bibnamefont{Mazin}},
  \bibinfo{author}{\bibfnamefont{M.~D.} \bibnamefont{Johannes}},
  \bibinfo{author}{\bibfnamefont{L.}~\bibnamefont{Boeri}}, \bibnamefont{and}
  \bibinfo{author}{\bibfnamefont{D.~S.} \bibnamefont{K.~Koepernik}},
  \bibinfo{journal}{Phys. Rev. B} \textbf{\bibinfo{volume}{78}},
  \bibinfo{pages}{085104} (\bibinfo{year}{2008}).

\bibitem[{\citenamefont{Raghu et~al.}(2008)\citenamefont{Raghu, Qi, Liu,
  Scalapino, and Zhang}}]{raghu08}
\bibinfo{author}{\bibfnamefont{S.}~\bibnamefont{Raghu}},
  \bibinfo{author}{\bibfnamefont{X.}~\bibnamefont{Qi}},
  \bibinfo{author}{\bibfnamefont{C.-X.} \bibnamefont{Liu}},
  \bibinfo{author}{\bibfnamefont{D.}~\bibnamefont{Scalapino}},
  \bibnamefont{and} \bibinfo{author}{\bibfnamefont{S.-C.} \bibnamefont{Zhang}},
  \bibinfo{journal}{Phys. Rev. B} \textbf{\bibinfo{volume}{77}},
  \bibinfo{pages}{220503} (\bibinfo{year}{2008}).

\bibitem[{\citenamefont{Chubukov et~al.}(2008)\citenamefont{Chubukov, Efremov,
  and Eremin}}]{chubukov08}
\bibinfo{author}{\bibfnamefont{A.}~\bibnamefont{Chubukov}},
  \bibinfo{author}{\bibfnamefont{D.}~\bibnamefont{Efremov}}, \bibnamefont{and}
  \bibinfo{author}{\bibfnamefont{I.}~\bibnamefont{Eremin}},
  \bibinfo{journal}{Phys. Rev. B} \textbf{\bibinfo{volume}{78}},
  \bibinfo{pages}{134512} (\bibinfo{year}{2008}).

\bibitem[{\citenamefont{Cvetkovic and Tesanovic}(2009)}]{cvetkovic09}
\bibinfo{author}{\bibfnamefont{V.}~\bibnamefont{Cvetkovic}} \bibnamefont{and}
  \bibinfo{author}{\bibfnamefont{Z.}~\bibnamefont{Tesanovic}},
  \bibinfo{journal}{Europhysics Lett.} \textbf{\bibinfo{volume}{85}},
  \bibinfo{pages}{37002} (\bibinfo{year}{2009}).

\bibitem[{\citenamefont{Qazilbash et~al.}(2009)\citenamefont{Qazilbash, Hamlin,
  Baumbach, Zhang, Singh, Maple, and Basov}}]{qazilbash09}
\bibinfo{author}{\bibfnamefont{M.}~\bibnamefont{Qazilbash}},
  \bibinfo{author}{\bibfnamefont{J.}~\bibnamefont{Hamlin}},
  \bibinfo{author}{\bibfnamefont{R.}~\bibnamefont{Baumbach}},
  \bibinfo{author}{\bibfnamefont{L.}~\bibnamefont{Zhang}},
  \bibinfo{author}{\bibfnamefont{D.}~\bibnamefont{Singh}},
  \bibinfo{author}{\bibfnamefont{M.}~\bibnamefont{Maple}}, \bibnamefont{and}
  \bibinfo{author}{\bibfnamefont{D.}~\bibnamefont{Basov}},
  \bibinfo{journal}{Electronic correlations in the iron pnictides}
  \textbf{\bibinfo{volume}{5}}, \bibinfo{pages}{647} (\bibinfo{year}{2009}).

\bibitem[{\citenamefont{Haule and Kotliar}(2009)}]{haule09}
\bibinfo{author}{\bibfnamefont{K.}~\bibnamefont{Haule}} \bibnamefont{and}
  \bibinfo{author}{\bibfnamefont{G.}~\bibnamefont{Kotliar}},
  \bibinfo{journal}{New Journal of Physics} \textbf{\bibinfo{volume}{11}},
  \bibinfo{pages}{025021} (\bibinfo{year}{2009}).

\bibitem[{\citenamefont{Ma et~al.}(2009)\citenamefont{Ma, Ji, Hu, Lu, and
  Xiang}}]{maPRL09}
\bibinfo{author}{\bibfnamefont{F.}~\bibnamefont{Ma}},
  \bibinfo{author}{\bibfnamefont{W.}~\bibnamefont{Ji}},
  \bibinfo{author}{\bibfnamefont{J.}~\bibnamefont{Hu}},
  \bibinfo{author}{\bibfnamefont{Z.-Y.} \bibnamefont{Lu}}, \bibnamefont{and}
  \bibinfo{author}{\bibfnamefont{T.}~\bibnamefont{Xiang}},
  \bibinfo{journal}{Phys. Rev. Lett.} \textbf{\bibinfo{volume}{102}},
  \bibinfo{pages}{177003} (\bibinfo{year}{2009}).

\bibitem[{\citenamefont{Moon et~al.}(2009)\citenamefont{Moon, Park, and
  Choi}}]{moonPRB09}
\bibinfo{author}{\bibfnamefont{C.-Y.} \bibnamefont{Moon}},
  \bibinfo{author}{\bibfnamefont{S.~Y.} \bibnamefont{Park}}, \bibnamefont{and}
  \bibinfo{author}{\bibfnamefont{H.~J.} \bibnamefont{Choi}},
  \bibinfo{journal}{Phys. Rev. B} \textbf{\bibinfo{volume}{80}},
  \bibinfo{pages}{054522} (\bibinfo{year}{2009}).

\bibitem[{\citenamefont{Moon and Choi}(2010)}]{moonPRL10}
\bibinfo{author}{\bibfnamefont{C.-Y.} \bibnamefont{Moon}} \bibnamefont{and}
  \bibinfo{author}{\bibfnamefont{H.~J.} \bibnamefont{Choi}},
  \bibinfo{journal}{Phys. Rev. Lett.} \textbf{\bibinfo{volume}{104}},
  \bibinfo{pages}{057003} (\bibinfo{year}{2010}).

\bibitem[{\citenamefont{Yan et~al.}()\citenamefont{Yan, Gao, Lu, and
  Xiang}}]{yan10}
\bibinfo{author}{\bibfnamefont{X.-W.} \bibnamefont{Yan}},
  \bibinfo{author}{\bibfnamefont{M.}~\bibnamefont{Gao}},
  \bibinfo{author}{\bibfnamefont{Z.-Y.} \bibnamefont{Lu}}, \bibnamefont{and}
  \bibinfo{author}{\bibfnamefont{T.}~\bibnamefont{Xiang}},
  \bibinfo{note}{arXiv:1012.5536}.

\bibitem[{\citenamefont{Zhao et~al.}(2009)\citenamefont{Zhao, Adroja, Yao,
  Bewley, Li, Wang, Wu, Chen, Hu, and Dai}}]{zhaonatphys09}
\bibinfo{author}{\bibfnamefont{J.}~\bibnamefont{Zhao}},
  \bibinfo{author}{\bibfnamefont{D.~T.} \bibnamefont{Adroja}},
  \bibinfo{author}{\bibfnamefont{D.-X.} \bibnamefont{Yao}},
  \bibinfo{author}{\bibfnamefont{R.}~\bibnamefont{Bewley}},
  \bibinfo{author}{\bibfnamefont{S.}~\bibnamefont{Li}},
  \bibinfo{author}{\bibfnamefont{X.~F.} \bibnamefont{Wang}},
  \bibinfo{author}{\bibfnamefont{G.}~\bibnamefont{Wu}},
  \bibinfo{author}{\bibfnamefont{X.~H.} \bibnamefont{Chen}},
  \bibinfo{author}{\bibfnamefont{J.}~\bibnamefont{Hu}}, \bibnamefont{and}
  \bibinfo{author}{\bibfnamefont{P.}~\bibnamefont{Dai}},
  \bibinfo{journal}{Nature Physics} \textbf{\bibinfo{volume}{5}},
  \bibinfo{pages}{555} (\bibinfo{year}{2009}).

\bibitem[{\citenamefont{Wysocki et~al.}(2011)\citenamefont{Wysocki,
  Belashchenko, and Antropov}}]{wysocki11}
\bibinfo{author}{\bibfnamefont{A.}~\bibnamefont{Wysocki}},
  \bibinfo{author}{\bibfnamefont{K.}~\bibnamefont{Belashchenko}},
  \bibnamefont{and} \bibinfo{author}{\bibfnamefont{V.~P.}
  \bibnamefont{Antropov}}, \bibinfo{journal}{Nature Physics}
  \textbf{\bibinfo{volume}{7}}, \bibinfo{pages}{485} (\bibinfo{year}{2011}).

\bibitem[{\citenamefont{Hu et~al.}()\citenamefont{Hu, Xu, Liu, Hao, and
  Wang}}]{strong-coupling-hu11}
\bibinfo{author}{\bibfnamefont{J.}~\bibnamefont{Hu}},
  \bibinfo{author}{\bibfnamefont{B.}~\bibnamefont{Xu}},
  \bibinfo{author}{\bibfnamefont{W.}~\bibnamefont{Liu}},
  \bibinfo{author}{\bibfnamefont{N.-N.} \bibnamefont{Hao}}, \bibnamefont{and}
  \bibinfo{author}{\bibfnamefont{Y.}~\bibnamefont{Wang}},
  \bibinfo{note}{arXiv:1106.5169}.

\bibitem[{\citenamefont{Yildirim}(2009)}]{yildirim09}
\bibinfo{author}{\bibfnamefont{T.}~\bibnamefont{Yildirim}},
  \bibinfo{journal}{Physica C} \textbf{\bibinfo{volume}{469}},
  \bibinfo{pages}{425} (\bibinfo{year}{2009}).

\bibitem[{\citenamefont{Xia et~al.}(2009)\citenamefont{Xia, Qian, Wray, Hsieh,
  Chen, Luo, Wang, and Hasan}}]{xiaPRL09}
\bibinfo{author}{\bibfnamefont{Y.}~\bibnamefont{Xia}},
  \bibinfo{author}{\bibfnamefont{D.}~\bibnamefont{Qian}},
  \bibinfo{author}{\bibfnamefont{L.}~\bibnamefont{Wray}},
  \bibinfo{author}{\bibfnamefont{D.}~\bibnamefont{Hsieh}},
  \bibinfo{author}{\bibfnamefont{G.~F.} \bibnamefont{Chen}},
  \bibinfo{author}{\bibfnamefont{J.~L.} \bibnamefont{Luo}},
  \bibinfo{author}{\bibfnamefont{N.~L.} \bibnamefont{Wang}}, \bibnamefont{and}
  \bibinfo{author}{\bibfnamefont{M.~Z.} \bibnamefont{Hasan}},
  \bibinfo{journal}{Phys. Rev. Lett.} \textbf{\bibinfo{volume}{103}},
  \bibinfo{pages}{037002} (\bibinfo{year}{2009}).

\bibitem[{\citenamefont{Hansmann et~al.}(2010)\citenamefont{Hansmann, Arita,
  Toschi, Sakai, Sangiovanni, and Held}}]{hansmannPRL10}
\bibinfo{author}{\bibfnamefont{P.}~\bibnamefont{Hansmann}},
  \bibinfo{author}{\bibfnamefont{R.}~\bibnamefont{Arita}},
  \bibinfo{author}{\bibfnamefont{A.}~\bibnamefont{Toschi}},
  \bibinfo{author}{\bibfnamefont{S.}~\bibnamefont{Sakai}},
  \bibinfo{author}{\bibfnamefont{G.}~\bibnamefont{Sangiovanni}},
  \bibnamefont{and} \bibinfo{author}{\bibfnamefont{K.}~\bibnamefont{Held}},
  \bibinfo{journal}{Phys. Rev. Lett.} \textbf{\bibinfo{volume}{104}},
  \bibinfo{pages}{197002} (\bibinfo{year}{2010}).

\bibitem[{\citenamefont{Yin et~al.}(2010{\natexlab{b}})\citenamefont{Yin,
  Haule, and Kotliar}}]{yin11}
\bibinfo{author}{\bibfnamefont{Z.~P.} \bibnamefont{Yin}},
  \bibinfo{author}{\bibfnamefont{K.}~\bibnamefont{Haule}}, \bibnamefont{and}
  \bibinfo{author}{\bibfnamefont{G.}~\bibnamefont{Kotliar}},
  \bibinfo{journal}{Nature Physics} \textbf{\bibinfo{volume}{7}},
  \bibinfo{pages}{294} (\bibinfo{year}{2010}{\natexlab{b}}).

\bibitem[{\citenamefont{Johannes and Mazin}(2009)}]{johannes09}
\bibinfo{author}{\bibfnamefont{M.}~\bibnamefont{Johannes}} \bibnamefont{and}
  \bibinfo{author}{\bibfnamefont{I.}~\bibnamefont{Mazin}},
  \bibinfo{journal}{Phys. Rev. B} \textbf{\bibinfo{volume}{79}},
  \bibinfo{pages}{220510R} (\bibinfo{year}{2009}).

\bibitem[{\citenamefont{Calder\'on et~al.}(2009)\citenamefont{Calder\'on,
  Valenzuela, and Bascones}}]{nosotrasprb09}
\bibinfo{author}{\bibfnamefont{M.~J.} \bibnamefont{Calder\'on}},
  \bibinfo{author}{\bibfnamefont{B.}~\bibnamefont{Valenzuela}},
  \bibnamefont{and} \bibinfo{author}{\bibfnamefont{E.}~\bibnamefont{Bascones}},
  \bibinfo{journal}{Phys. Rev. B} \textbf{\bibinfo{volume}{80}},
  \bibinfo{pages}{094531} (\bibinfo{year}{2009}).

\bibitem[{\citenamefont{Singh et~al.}(2009)\citenamefont{Singh, Green, Huang,
  Kreyssig, McQueeney, Johnston, and Goldman}}]{singh_mn09}
\bibinfo{author}{\bibfnamefont{Y.}~\bibnamefont{Singh}},
  \bibinfo{author}{\bibfnamefont{M.}~\bibnamefont{Green}},
  \bibinfo{author}{\bibfnamefont{Q.}~\bibnamefont{Huang}},
  \bibinfo{author}{\bibfnamefont{A.}~\bibnamefont{Kreyssig}},
  \bibinfo{author}{\bibfnamefont{R.}~\bibnamefont{McQueeney}},
  \bibinfo{author}{\bibfnamefont{D.}~\bibnamefont{Johnston}}, \bibnamefont{and}
  \bibinfo{author}{\bibfnamefont{A.}~\bibnamefont{Goldman}},
  \bibinfo{journal}{Phys. Rev. B} \textbf{\bibinfo{volume}{80}},
  \bibinfo{pages}{100403(R)} (\bibinfo{year}{2009}).

\bibitem[{\citenamefont{Yanagi et~al.}(2008)\citenamefont{Yanagi, Kawamura,
  Kamiya, Kamihara, Hirano, Nakamura, Osawa, and Hosono}}]{yanagi08}
\bibinfo{author}{\bibfnamefont{H.}~\bibnamefont{Yanagi}},
  \bibinfo{author}{\bibfnamefont{R.}~\bibnamefont{Kawamura}},
  \bibinfo{author}{\bibfnamefont{T.}~\bibnamefont{Kamiya}},
  \bibinfo{author}{\bibfnamefont{Y.}~\bibnamefont{Kamihara}},
  \bibinfo{author}{\bibfnamefont{M.}~\bibnamefont{Hirano}},
  \bibinfo{author}{\bibfnamefont{T.}~\bibnamefont{Nakamura}},
  \bibinfo{author}{\bibfnamefont{H.}~\bibnamefont{Osawa}}, \bibnamefont{and}
  \bibinfo{author}{\bibfnamefont{H.}~\bibnamefont{Hosono}},
  \bibinfo{journal}{Physical Review B} \textbf{\bibinfo{volume}{77}},
  \bibinfo{pages}{224431} (\bibinfo{year}{2008}).

\bibitem[{\citenamefont{Castellani et~al.}(1978)\citenamefont{Castellani,
  Natoli, and Ranninger}}]{castellani78}
\bibinfo{author}{\bibfnamefont{C.}~\bibnamefont{Castellani}},
  \bibinfo{author}{\bibfnamefont{C.~R.} \bibnamefont{Natoli}},
  \bibnamefont{and}
  \bibinfo{author}{\bibfnamefont{J.}~\bibnamefont{Ranninger}},
  \bibinfo{journal}{Phys. Rev. B} \textbf{\bibinfo{volume}{18}},
  \bibinfo{pages}{4945} (\bibinfo{year}{1978}).

\bibitem[{\citenamefont{Xu et~al.}(2008)\citenamefont{Xu, M\"uller, and
  Sachdev}}]{sachdev08}
\bibinfo{author}{\bibfnamefont{C.}~\bibnamefont{Xu}},
  \bibinfo{author}{\bibfnamefont{M.}~\bibnamefont{M\"uller}}, \bibnamefont{and}
  \bibinfo{author}{\bibfnamefont{S.}~\bibnamefont{Sachdev}},
  \bibinfo{journal}{Phys. Rev. B} \textbf{\bibinfo{volume}{78}},
  \bibinfo{pages}{020501} (\bibinfo{year}{2008}).

\bibitem[{\citenamefont{Fang et~al.}(2008)\citenamefont{Fang, Yao, Tsai, Hu,
  and Kivelson}}]{kivelson08}
\bibinfo{author}{\bibfnamefont{C.}~\bibnamefont{Fang}},
  \bibinfo{author}{\bibfnamefont{H.}~\bibnamefont{Yao}},
  \bibinfo{author}{\bibfnamefont{W.-F.} \bibnamefont{Tsai}},
  \bibinfo{author}{\bibfnamefont{J.}~\bibnamefont{Hu}}, \bibnamefont{and}
  \bibinfo{author}{\bibfnamefont{S.~A.} \bibnamefont{Kivelson}},
  \bibinfo{journal}{Phys. Rev. B} \textbf{\bibinfo{volume}{77}},
  \bibinfo{pages}{224509} (\bibinfo{year}{2008}).

\bibitem[{\citenamefont{Bascones et~al.}(2010)\citenamefont{Bascones,
  Calder\'on, and Valenzuela}}]{nosotrasprl10}
\bibinfo{author}{\bibfnamefont{E.}~\bibnamefont{Bascones}},
  \bibinfo{author}{\bibfnamefont{M.~J.} \bibnamefont{Calder\'on}},
  \bibnamefont{and}
  \bibinfo{author}{\bibfnamefont{B.}~\bibnamefont{Valenzuela}},
  \bibinfo{journal}{Phys. Rev. Lett.} \textbf{\bibinfo{volume}{104}},
  \bibinfo{pages}{227201} (\bibinfo{year}{2010}).

\bibitem[{\citenamefont{Valenzuela et~al.}(2010)\citenamefont{Valenzuela,
  Bascones, and Calder\'on}}]{nosotrasprl10_2}
\bibinfo{author}{\bibfnamefont{B.}~\bibnamefont{Valenzuela}},
  \bibinfo{author}{\bibfnamefont{E.}~\bibnamefont{Bascones}}, \bibnamefont{and}
  \bibinfo{author}{\bibfnamefont{M.~J.} \bibnamefont{Calder\'on}},
  \bibinfo{journal}{Phys. Rev. Lett.} \textbf{\bibinfo{volume}{105}},
  \bibinfo{pages}{207202} (\bibinfo{year}{2010}).

\bibitem[{\citenamefont{Laad and Craco}()}]{laad2011}
\bibinfo{author}{\bibfnamefont{M.}~\bibnamefont{Laad}} \bibnamefont{and}
  \bibinfo{author}{\bibfnamefont{L.}~\bibnamefont{Craco}},
  \bibinfo{note}{arxiv:1010.2940}.

\bibitem[{\citenamefont{Cricchio et~al.}(2009)\citenamefont{Cricchio, Granas,
  and Nordstrom}}]{cricchio09}
\bibinfo{author}{\bibfnamefont{F.}~\bibnamefont{Cricchio}},
  \bibinfo{author}{\bibfnamefont{O.}~\bibnamefont{Granas}}, \bibnamefont{and}
  \bibinfo{author}{\bibfnamefont{L.}~\bibnamefont{Nordstrom}},
  \bibinfo{journal}{Phys. Rev. B} \textbf{\bibinfo{volume}{81}},
  \bibinfo{pages}{140403} (\bibinfo{year}{2009}).

\bibitem[{\citenamefont{Liu}(2011)}]{liu11}
\bibinfo{author}{\bibfnamefont{G.-Q.} \bibnamefont{Liu}},
  \bibinfo{journal}{arXiv:1105.5412v1}  (\bibinfo{year}{2011}).

\bibitem[{\citenamefont{Lee et~al.}(2008)\citenamefont{Lee, Iyo, Eisaki, Kito,
  Fernandez-Diaz, Ito, Kihou, Matsuhata, Braden, and Yamada}}]{iyo08}
\bibinfo{author}{\bibfnamefont{C.~H.} \bibnamefont{Lee}},
  \bibinfo{author}{\bibfnamefont{A.}~\bibnamefont{Iyo}},
  \bibinfo{author}{\bibfnamefont{H.}~\bibnamefont{Eisaki}},
  \bibinfo{author}{\bibfnamefont{H.}~\bibnamefont{Kito}},
  \bibinfo{author}{\bibfnamefont{M.~T.} \bibnamefont{Fernandez-Diaz}},
  \bibinfo{author}{\bibfnamefont{T.}~\bibnamefont{Ito}},
  \bibinfo{author}{\bibfnamefont{K.}~\bibnamefont{Kihou}},
  \bibinfo{author}{\bibfnamefont{H.}~\bibnamefont{Matsuhata}},
  \bibinfo{author}{\bibfnamefont{M.}~\bibnamefont{Braden}}, \bibnamefont{and}
  \bibinfo{author}{\bibfnamefont{K.}~\bibnamefont{Yamada}},
  \bibinfo{journal}{J. Phys. Soc. Jpn.} \textbf{\bibinfo{volume}{77}},
  \bibinfo{pages}{083704} (\bibinfo{year}{2008}).

\bibitem[{\citenamefont{Kuroki et~al.}(2009)\citenamefont{Kuroki, Usui, Onari,
  Arita, and Aoki}}]{kuroki09-2}
\bibinfo{author}{\bibfnamefont{K.}~\bibnamefont{Kuroki}},
  \bibinfo{author}{\bibfnamefont{H.}~\bibnamefont{Usui}},
  \bibinfo{author}{\bibfnamefont{S.}~\bibnamefont{Onari}},
  \bibinfo{author}{\bibfnamefont{R.}~\bibnamefont{Arita}}, \bibnamefont{and}
  \bibinfo{author}{\bibfnamefont{H.}~\bibnamefont{Aoki}},
  \bibinfo{journal}{Phys. Rev. B} \textbf{\bibinfo{volume}{79}},
  \bibinfo{pages}{224511} (\bibinfo{year}{2009}).

\bibitem[{not()}]{note-cf}
\bibinfo{note}{The effect of the Fe-As-Fe angle on the crystal field has been
  neglected.}

\bibitem[{\citenamefont{Misawa et~al.}(2012)\citenamefont{Misawa, Nakamura, and
  Imada}}]{imadaPRL2012}
\bibinfo{author}{\bibfnamefont{T.}~\bibnamefont{Misawa}},
  \bibinfo{author}{\bibfnamefont{K.}~\bibnamefont{Nakamura}}, \bibnamefont{and}
  \bibinfo{author}{\bibfnamefont{M.}~\bibnamefont{Imada}},
  \bibinfo{journal}{Phys. Rev. Lett.} \textbf{\bibinfo{volume}{108}},
  \bibinfo{pages}{177007} (\bibinfo{year}{2012}).

\end{thebibliography}
\end{document}